\documentstyle[aps,preprint]{revtex}
\tightenlines
\begin{document}
\preprint{}
\title{Effects of R-parity violation on
CP asymmetries\\ in $\Lambda_b \to p \pi$ decay}
\author{Rukmani Mohanta}
\address{Physics Department, Panjab University, 
Chandigarh - 160014, India}
\maketitle
\begin{abstract}

We have studied new CP violating effects in $\Lambda_b \to p \pi$
decay mode, that can arise in Minimal Supersymmetric Standard Model
with R-parity violation. We have estimated how much R-parity violation
modifies the Standard Model predictions for CP asymmetries within the
present bounds. We found that in the R-parity violating model,
the rate asymmetry ($a_{cp}$) is suppressed (about 10 times)
and the asymmetry parameter $A(\alpha)$ is enhanced 
(approximately ${\cal O}(10^2)$ times) with respect to
the SM predictions.

\end{abstract}
\pacs{PACS Nos. : 11.30.Er, 12.60.Jv., 13.30.Eg}

One of the most important objects of the upcoming experiments at 
$B$ factories is to search for CP violation in as many $B$ decay
modes as possible so as to establish the pattern of CP violation
among various  $B$ decays \cite{ref1}. This then may allow
for an experimental test not only of the Standard Model (SM) 
Cabibbo-Kobayashi-Maskawa (CKM) paradigm of CP violation,
but also many extensions of the SM that often contain new sources of CP
violation. It is well known that CP violating $B$ decays might
constitute an important hunting ground for new physics. This is 
particularly so since many CP violating asymmetries related to $B$
decays are predicted to be small in SM, are likely to be measured
with high precision in the upcoming $B$ factories. Measurements
larger than the SM predictions would definitely signal the presence
of new physics. It is also interesting to study CP violation in bottom
baryon system in order to find the physical channels which may have
large CP asymmetries, even though the branching ratios for such
processes are usually smaller than those for the corresponding
bottom mesons. Recently some data on the bottom baryon $\Lambda_b$
have appeared. For instance, OPAL has measured its lifetime and 
the production branching ratio for the inclusivs semileptonic decay
$\Lambda_b \to \Lambda l^- \bar {\nu} X$ \cite{ref2}. Furthermore,
mesurements of the nonleptonic decay $\Lambda_b \to \Lambda J/\psi$
has also been reported \cite{ref3}. Certainly we expect more data
in the future in the bottom baryon sector. In this paper we intend to
study CP violation in the nonleptonic $\Lambda_b \to p \pi$ decay in
the Minimal Supersymmetric Standard Model (MSSM) with $R$-parity
violation \cite{ref4}. The MSSM has been widely considered as a
leading candidate for new physics beyond SM. 
In supersymmetric theories `$R$-parity' is a discrete
symmetry under which all standard model particles are even while
their superpartner are odd. It is defined as $R = (-1)^{(3B+L+2S)}$,
where $S$ is the spin, $B$ is the baryon number and $L$ is the lepton
number of the particle. An exact $R$-parity implies that superparticles
could be produced in pairs and the lightest supersymmetric particle
(LSP) is stable. However, $B$ and $L$ conservations are not ensured by
gauge invariance and therefore it is worhwhile to investigate what
happens to the CP asymmetries when  $R$-parity is violated.

The most general Lorentz-invariant amplitude
for the decay $\Lambda_b \to p \pi^- $ can be written as \cite{ref5}
\begin{equation}
i \bar u_{p}(p_f)(a+b \gamma_5)u_{\Lambda_b}(p_i)
\end{equation}
The corresponding matrix element for $\bar \Lambda_b \to 
\bar p \pi^+$ is then
\begin{equation}
i \bar v_{\bar p}(p_f)(-a^*+b^* \gamma_5)v_{\bar \Lambda_b}(p_i)
\end{equation}
It is convenient to express the transition amplitude 
in terms of S-wave (parity violating) and P-wave (parity conserving) 
amplitudes S and P as

\begin{equation}
S+P \sigma \cdot \hat{\bf q}
\end{equation}
where ${\bf q}$ is the proton momentum in the rest frame 
of $\Lambda_b$ baryon and the amplitudes S and P are :

\begin{equation}
S=a \sqrt{\frac{\{(m_{\Lambda_b}+m_p)^2-m_{\pi}^2 \}}
{16 \pi m_{\Lambda_b}^2}}~~~~
P=b \sqrt{\frac{\{(m_{\Lambda_b}-m_p)^2-m_{\pi}^2 \}}
{16 \pi m_{\Lambda_b}^2}}
\end{equation}
The experimental observables are the total decay rate 
$\Gamma$ and the decay parameters $\alpha$, $\beta$ and 
$\gamma$ which govern the decay-angular
distribution and the polarization of the final baryon. The decay 
rate is given as
\begin{equation}
\Gamma=2 |{\bf q}|\{|S|^2+|P|^2\}
\end{equation}
and the dominant asymmetry parameter ($\alpha$) is given as
\begin{equation}
\alpha=\frac{2 Re (S^*P)}{\{|S|^2+|P|^2\}}\;
\end{equation}
Similar observables for the antihyperon decays are $\bar \Gamma$
and $\bar \alpha$ are given as
\begin{equation}
\bar \Gamma=2 |{\bf q}|\{|\bar S|^2+| \bar P|^2\}\;,
~~~~~~
\bar \alpha=\frac{2 Re (\bar S^*\bar P)}{\{| \bar S|^2+|\bar P|^2\}}
\end{equation}

For $\Lambda_b \to p \pi^-$ decay the CP violating
rate asymmetry in partial decay rate ($a_{cp}$) and  
aymmetry parameter ($A(\alpha)$)
are defined as follows,
\begin{equation}
a_{cp} =\frac{\Gamma (\Lambda_b \to p \pi^-) - \Gamma 
(\bar \Lambda_b \to \bar p \pi^+)}
{\Gamma (\Lambda_b \to p \pi^-) + \Gamma 
(\bar \Lambda_b \to \bar p \pi^+)}\;,
\end{equation}
\begin{equation}
A(\alpha)=\frac{\alpha + \bar \alpha}{\alpha - \bar \alpha}\;,
\end{equation}
A nonzero value for $a_{cp}$
and $A(\alpha)$ will signal CP violation. The existence of such CP 
asymmetries require the interference of two decay amplitudes with
different weak and strong phase differences. The weak phase difference
arises from the superposition of various penguin contributions and 
the usual tree diagrams while the strong phases are induced by
final state interactions (FSI). At the quark level, the strong phase
diffences arise through the absorptive parts of penguin diagrams
(hard final state interactions) \cite{ref6} and nonperturbatively
(soft final state interactions) \cite{ref7}. In the absence of an
argument that the parton-hadron duality should hold in exclusive 
processes, one can not exclude that the weak transition matrix elements
receive phases originating from soft FSI. However the effects of soft
FSI are extremely difficult to quantify. In the absence of a reliable
theoretical calculation for soft FSI, we make the usual approximation
of retaining the absorptive part from quark level calculation (hard FSI)
for strong phase differences in our analysis.

We shall first consider the SM contributions to the transition
amplitude. The effective Hamiltonian ${\cal H}_{eff} $ for the decay process 
$\Lambda_b \to p \pi^- $ is given as
\begin{equation}
{\cal H}_{eff}= \frac{G_F}{\sqrt 2} \biggr\{V_{ub}V_{ud}^*
[c_1(\mu) O_1^u(\mu)+c_2(\mu) O_2^u(\mu)]-
V_{tb}V_{td}^*\sum_{i=3}^{10} c_i(\mu) O_i(\mu)\biggr\}
+{\rm h.c.}\;,
\end{equation}
where $O_{1,2}$ are the tree level current-current operators, $O_{3-6}$
are the QCD and $O_{7-10}$ are the electroweak penguin operators
which are explicitly given in Ref. \cite{ref8}, $c_i$'s are the Wilson 
coefficients, which take care of the short-distance QCD corrections, 
are scheme and scale dependent. These unphysical dependences are
cancelled by the corresponding scheme and scale dependences of the
matrix elements of the operators. However, in the factorization
approximation, the hadronic matrix elements are written in terms of
form factors and decay constants, which are scheme and scale
independent. So to achieve the cancellation, the various one loop 
corrections are absorbed into the effective Wilson coefficients
$c_i^{eff}$, which are scheme and scale independent. 
The values of the effective Wilson coefficients for $b \to d $ transitions
are explicitly evaluated in Ref. \cite{ref8} as :

\begin{eqnarray}
&&c_1^{eff}=1.168~~~~ c_2^{eff} =-0.365~~~~c_3^{eff}=0.0224+i0.0038
~~~~c_4^{eff}=-(0.0454+i0.0115)\nonumber\\
&&c_5^{eff}=0.0131+i0.0038
~~~c_6^{eff}=-(0.0475+i0.0115)
~~~c_7^{eff}/\alpha=-(0.0294+i0.0329)\nonumber\\
&&c_8^{eff}/\alpha=0.055~~~~c_9^{eff}/\alpha=-(1.426+i0.0329)
~~~~~c_{10}^{eff}/\alpha=0.48
\end{eqnarray}
These one loop corrections (to get $c_i^{eff}$'s ) result in imaginary 
parts for ($c_i^{eff}$'s) due to virtual quarks going on their mass 
shell. 

The matrix elements of the operators can be calculated using the
factorization approximation. In this approximation the hadronic
matrix elements of the four quark operators $(\bar d b)_{V-A}
(\bar u d)_{V-A}$ split into products of matrix elements one
involving pion decay constant and the other dealt the baryonic
form factors. The matrix elements of the $(V-A)(V+A)$ i.e., ($ O_6$
and $O_8$) operators can be obtained by Fierz reordering and using
the Dirac equation as,

\begin{equation}
\langle p \pi |O_6 |\Lambda_b \rangle
=-2 \sum_q \langle  \pi |\bar d(1+\gamma_5) q |0 \rangle
\langle p |\bar q (1-\gamma_5) b |\Lambda_b \rangle
\end{equation}
Using the Dirac equation the matrix element can be rewritten as
\begin{equation}
\langle p \pi |O_6 |\Lambda_b \rangle=\biggr[R_1
\langle p |V_\mu |\Lambda_b \rangle
-R_2\langle p |A_\mu |\Lambda_b \rangle \biggr]
\langle \pi | A_{\mu}|0\rangle\;,
\end{equation}
with
\begin{equation}
R_1=\frac{2 m_{\pi}^2}{(m_b-m_u)(m_d+m_u)}\;,
~~~~~~~~R_2=\frac{2 m_{\pi}^2}{(m_b+m_u)(m_d+m_u)}\;,
\end{equation}
where the quark masses are the current quark masses. The same
relation works for $O_8$.

Thus under the factorization approximation the baryon decay
amplitude is governed by a decay constant and baryonic transition
form factors.
The general expression for
the baryon transition is given as
\begin{eqnarray}
\langle p(p_f)|V_{\mu}-A_{\mu}|\Lambda_b(p_i) \rangle 
&=&\bar u_p(p_f) \biggr\{f_1(q^2) \gamma_{\mu} 
+i f_2 (q^2) \sigma_{\mu \nu}
q^{\nu}+f_3 (q^2) q_\mu \nonumber\\
&-&[g_1(q^2) \gamma_{\mu} +i g_2 (q^2) \sigma_{\mu \nu}
q^{\nu}+g_3 (q^2) q_\mu]\gamma_5 \biggr\}u_{\Lambda_b}(p_i)\;,
\end{eqnarray}
where $q=p_i-p_f$.
The values of the form factors at maximum momentum transfer
are evaluated in nonrelativistic quark model and their $q^2$
dependence are determined using the pole dominance
model \cite{ref9} with values as,
\begin{equation}
f_1(m_\pi^2)=0.043~~~m_if_3(m_\pi^2)=-0.009~~~g_1(m_\pi^2)=0.092~~~
m_ig_3(m_\pi^2)=-0.047\;,
\end{equation}
where the particle masses are taken from \cite{ref10}.

Hence one obtains the amplitude for the decay mode
$\Lambda_b \to p \pi^-$ as
(where the factor $G_F/\sqrt 2 $ is suppressed)
\begin{eqnarray}
A(\Lambda_b  \to   p \pi^- )&=&i f_{\pi} \bar u_p(p_f)
\biggr[\biggr\{\lambda_u  \left (a_1+a_4+a_{10}+(a_6+a_8)R_1 \right )
+\lambda_c  \left (a_4+a_{10}+(a_6+a_8)
R_1 \right )\biggr\}\nonumber\\
&\times &
\left (f_1(m_{\pi}^2)(m_i-m_f)+f_3(m_{\pi}^2) 
m_{\pi}^2 \right )\nonumber\\
&+&\biggr\{\lambda_u  \left (a_1+a_4+a_{10}+
(a_6+a_8)R_2 \right )
+\lambda_c  \left (a_4+a_{10}+(a_6+a_8)
R_2 \right )\biggr\}\nonumber\\
&\times &
\left (g_1(m_{\pi}^2)(m_i+m_f)
-g_3(m_{\pi}^2) m_{\pi}^2 \right )\gamma_5\biggr]u_{\Lambda_b}(p_i)\;,
\end{eqnarray}
where $m_i$ and $m_f$ are the masses of the initial and final baryons
respectively.
The coefficients $a_1,~a_2 \cdots~a_{10} $ are combinations of the
effective Wilson coefficients given as
\begin{equation}
a_{2i -1}=c_{2i-1}^{eff}+\frac{1}{(N_c)}c_{2i}^{eff}~~~
a_{2i}=c_{2i}^{eff}+\frac{1}{(N_c)}c_{2i-1}^{eff}~~~~
i=1,2 \cdots 5\;,
\end{equation}
where $N_c$ is the number of colors, taken to be $N_c=3$ 
for naive factorization. Thus one obtains the S and P-wave amplitudes
using eqns. (1), (4) and (17), in units of $(10^{-9})$ as

\begin{eqnarray}
&&S=  \lambda_u(34.603-0.7115 i )
-\lambda_c(2.782+0.7115 i)\nonumber\\
&&P=  \lambda_u(74.056-1.521 i )
-\lambda_c(5.95+1.521 i)\;,
\end{eqnarray}
with $\lambda_i = V_{ib} V_{id}$.
Now we shall proceed to evaluate the R-parity violating ($\not\!{R}_{p}$)
amplitude. In the 
MSSM the most general R-parity violating superpotential is given as
\begin{equation}
W_{\not\!{R}_{p}}=\lambda_{ijk}L_i L_j E_k^c +\lambda_{ijk}^\prime
L_i Q_j D_k^c + \lambda_{ijk}^{\prime \prime} U_i^c D_j^c D_k^c\;,
\label{eq:mat1}
\end{equation}
where $i,j,k $ are the generation indices and we assume that possible
bilinear terms $\mu_i L_i H_2 $ can be rotated away. $L_i $ and $Q_i$
are the $SU(2) $-doublets for lepton and quark superfields and $E_i^c$, 
$U_i^c$ and $D_i^c$ are the singlet superfields. $\lambda_{ijk}$ and
$\lambda_{ijk}^{\prime \prime}$ are antisymmetric under the interchange
of the first two and last two indices. The first two terms violate
lepton numbers where as the last term violates baryon number. For our
purpose we will consider only the lepton number violation contributions.
As the $\lambda $ type couplings do not contribute to the nonleptonic decays
we obtain from eqn. (\ref{eq:mat1}) the following effective Hamiltonian
due to the exchange of sleptons as
\begin{equation}
{\cal H}_{\not\!{R}_{p}}^{eff}=\sum_{n,p,q=1}^3 \frac{\lambda_{npi}^\prime
\lambda_{nql}^{\prime *}}{M_{\tilde l_n}^2}V_{kq} V_{jp}^*
(\bar d_i P_L u_j)(\bar u_k P_R d_l)
\end{equation}
with $P_{L,R}=(1 \mp \gamma_5)/2$.
From the above effective Hamiltonian we calculate the amplitude 
${\cal A}_{\not\!{R}_{p}}(\Lambda_b \to p \pi)$ using the factorization
approximation. The matrix elements of the $(S-P)(S+P) $ operators
are obtained using the Dirac equation of motion. Assuming $V_{CKM}$ 
is given by only down-type quark sector we obtain the dominant
transition amplitude to be
\begin{eqnarray}
{\cal A}_{\not\!{R}_{p}}^{eff}&=&\sum_{n,=2,3} \frac{\lambda_{npi}^\prime
\lambda_{nql}^{\prime *}}{M_{\tilde l_n}^2}V_{11} V_{11}^*~
\times ~i f_{\pi} \bar u(p_f)
\biggr[R_1\left (f_1(m_{\pi}^2)(m_i-m_f)+f_3(m_{\pi}^2) 
m_{\pi}^2 \right )\nonumber\\
&+&R_2\left (g_1(m_{\pi}^2)(m_i+m_f)
-g_3(m_{\pi}^2) m_{\pi}^2 \right )\gamma_5\biggr]u_{\Lambda_b}(p_i)\;.
\end{eqnarray}
Now considering the slepton mass to be 100 GeV, the present bounds
on $\lambda_{ijk}^\prime $ are \cite{ref11}
\begin{equation}
\lambda_{211}^\prime <0.09~~~~~\lambda_{213}^\prime <0.09
~~~~~\lambda_{311}^\prime <0.16~~~~\lambda_{313}^\prime <0.16
\end{equation}
we obtain the S and P-wave $R_{\not\!{R}_{p}}$ amplitudes to be
\begin{equation}
S_{\not\!{R}_{p}}<1.626 \times 10^{-9}~~~~P_{\not\!{R}_{p}}
<3.474 \times 10^{-9}
\end{equation}
After obtaining the transition amplitude in SM and $R_{\not\!{R}_{p}}$
model we now proceed to estimate the CP asymmetries. The parity
violating (S wave) and parity conserving (P wave) amplitudes can
be explicitly written as
\begin{eqnarray}
&&S =\lambda_u S_u+ \lambda_c S_c+S_{\not\!{R}_{p}}
\nonumber\\
&&P =\lambda_u P_u+ \lambda_c P_c+P_{\not\!{R}_{p}}
\end{eqnarray}
where $\lambda_i = V_{ib} V_{id}$, are the product of CKM matrix elements
which contain the weak phases. The 
strong phases which arise from the perturbative penguin 
diagrams at one loop level, are contained in $S_{u/c}$ and $P_{u/c}$
i.e.,  $S_u=|S_u|e^{i \delta_u}$ etc. 
The corresponding quantities for the antihyperon
decays are given as

\begin{eqnarray}
&&\bar S =-\left (\lambda_u^* S_u+ \lambda_c^* S_c
+S_{\not\!{R}_{p}}\right )
\nonumber\\
&&\bar P =\lambda_u^* P_u + 
\lambda_c^* P_c+P_{\not\!{R}_{p}}
\end{eqnarray}

Thus the CP violating rate asymmetry is given as,
\begin{equation}
a_{cp} =\frac{2 \biggr[Im(\lambda_u \lambda_c^*)\left ( Im(S_u S_c^*)
+Im(P_u P_c^*)\right )+Im(\lambda_u)[S_{\not\!{R}_{p}}Im(S_u)
+P_{\not\!{R}_{p}}Im(P_u)]\biggr]}{A}
\end{equation}
where
\begin{eqnarray}
A&=&\biggr[|\lambda_u|^2(|S_u|^2+|P_u|^2) +|\lambda_c |^2
(|S_c|^2+|P_c|^2) + (|S_{\not\!{R}_{p}}|^2+|P_{\not\!{R}_{p}}|^2)\nonumber\\
&+&2 Re(\lambda_u \lambda_c^*)
\left (Re(S_u S_c^*)+Re(P_u P_c^*)\right )+2\sum_{i=u,c}
Re(\lambda_i)(S_{\not\!{R}_{p}}Re(S_i)+P_{\not\!{R}_{p}}Re(P_i))\biggr]
\end{eqnarray}
Using the Wolfenstein parametrization for CKM matrix elements with values 
$A=0.815$, $ \lambda =0.2205 $, $\rho=0.175$ and $\eta=0.37 $, we obtain
the branching ratio and CP violating observables in RPV model 
using eqns. (5), (9) and (27) as
\begin{eqnarray}
&& Br(\Lambda_b \to p \pi) < 1.6 \times 10^{-4} \nonumber\\
&& a_{cp} \simeq 0.3\%\nonumber\\
&& A(\alpha) \simeq 8.8 \times 10^{-3}\label{eq:mat11}
\end{eqnarray}
The corresponding quantities in the SM ($S_{\not\!{R}_{p}}=0$
and $P_{\not\!{R}_{p}}=0$) are given as
\begin{eqnarray}
&& Br(\Lambda_b \to p \pi) = 0.9 \times 10^{-6} \nonumber\\
&& a_{cp} = 8.3\%\nonumber\\
&& A(\alpha) = 2.3 \times 10^{-5}\label{eq:mat12}
\end{eqnarray}
It can be seen from eqns. (\ref{eq:mat11}) and (\ref{eq:mat12})
that the effects of R-parity and lepton number violating couplings 
significantly modify the SM results of the branching ratio and
CP asymmetry parameters for the decay mode $\Lambda_b \to p\pi$.
The branching ratio and the asymmetry parameter ($A(\alpha)$)
in RPV model are approximately ${\cal O}(10^2)$ times larger than the
SM contributions whereas the rate asymmetry $a_{cp}$ is nearly 10 times 
smaller than the SM result.

To summarize, in this work we have studied the effects of R-parity 
violating couplings on the direct CP asymmetry parameters in $\Lambda_b
\to p \pi$ decay mode. Assuming factorization, we have used the
nonrelativistic quark model to evaluate the form factors at maximum 
momentum transfer ($q_m^2$) and the extrapolation of the form factors from 
$q_m^2$ to the required $q^2$ value is done by using the pole 
dominance. Although there are significant uncertainties in our
estimates as we have used the factorization approximation
to evaluate the matrix elements of the four-quark current
operators and taken all the R-parity violating couplings to be
real, it is probably safe to say that the asymmetry parameter
$A(\alpha)$ in $\Lambda_b \to p \pi $ decay is significantly
larger than the corresponding asymmetry in the Standard Model.

The author would like to thank Professor 
M. P. Khanna and Dr. A. K. Giri
for many useful discussions and also to CSIR, 
Govt. of India, for financial support.


\begin{thebibliography}{99}

\bibitem{ref1} For a recent review, see for example, {\it The
BaBar Physics Book}, ed. P. Harrison and H. Quinn, SLAC-R-504,
(1998); Y. Nir, hep-ph/9911321.
\bibitem{ref2} OPAL Collaboration, R. Akers et al., Z. Phys.
{\bf C 69}, 195 (1996); Phys. Lett. {\bf B 353}, 402 (1995); OPAL
Collaboration, K. Ackerstaff et al, Phys. Lett. {\bf B 426}, 161 (1998).
\bibitem{ref3} CDF Collaboration, F. Abe et al, Phys. Rev. 
{\bf D 55}, 1142 (1997): UA1 Collaboration, C. Albarjar et al., 
Phys. Lett. {\bf B 273}, 540 (1991): CDF Collaboration, 
F. Abe et al., Phys. Rev. {\bf D 47}, 2639 (1993); 
P. Abreu et al., Phys. Lett. {\bf B 374}, 351 (1996).
\bibitem{ref4} C. S. Aulakh and R. N. Mohapatra, Phys. Lett.
{\bf B 119}, 136 (1982); F. Zwirner, Phys. Lett. {\bf 132},
103 (1983); I.-H. Lee, Nucl. Phys. {\bf B 246}, 120 (1984);
J. Ellis et al, Phys. Lett. {\bf B 150}, 142 (1985); G. G. Ross
and J. W. F. Valle, Phys. Lett. {\bf 151}, 375 (1985); S. Dawson,
Nucl. Phys. {\bf B 261}, 295 (1985); R. Barbieri, A. Masiero,
Nucl. Phys. {\bf B 267}, 679 (1986): S. Dimopoulos and
L. Hall, Phys. Lett. {\bf B 207}, 210 (1987).
\bibitem{ref5} J. F. Donoghue and S. Pakvasa, 
Phys. Rev. Lett {\bf 55}, 162 (1985); J. F. Donoghue, X-G. He 
and S. Pakvasa, Phys. Rev. {\bf D 34}, 833 (1986); 
D. Chang, X-G. He and S. Pakvasa, Phys. Rev. Lett.
{\bf 74}, 3927 (1995); X-G. He, H. Steger and G. Valencia, Phys.
Lett. {\bf B 272}, 411 (1991); 
X-G. He and G. Valencia, Phys. Rev. {\bf D 52}, 5257 (1995); 
J. Tandean and G. Valencia, Phys. Lett. {\bf 451}, 382 (1999).
\bibitem{ref6} M. Bander, D. Silverman and A. Soni, Phys. Rev.
Lett. {\bf 43}, 242 (1979); J. M. G\'erard and W. S. Hou, Phys.
Rev. {\bf D 43}, 2909 (1991).
\bibitem{ref7} J. F. Donoghue, E. Golowich, A. A. Petrov and
J. M. Soares, Phys. Rev. Lett. {\bf 77}, 2178 (1996).
\bibitem{ref8} Y. H. Chen, H. Y. Cheng, B. Tseng and K. C. Yang,
Phys. Rev. {\bf D 60}, 094014 (1999).
\bibitem{ref9} H. Y. Cheng and B. Tseng, Phys. 
Rev. {\bf D 53}, 1457 (1996).
\bibitem{ref10} C. Caso et al, Euro. Phys. J. {\bf C
1}, 1 (1998).
\bibitem{ref11} H. Dreiner, hep-ph/9707435.

\end{thebibliography}
\end{document}